\documentclass[12pt]{iopart}
\usepackage{iopams}  
\usepackage{epsf}  
\usepackage[dvips]{graphics}

\newcommand{\beq}{\begin{equation}}
\newcommand{\eeq}{\end{equation}}
\newcommand{\beqa}{\begin{eqnarray}}
\newcommand{\eeqa}{\end{eqnarray}}

\newcommand{\xiv}{{\vec{\xi}}}
\newcommand{\Rv}{{\vec{R}}}
\newcommand{\Ocal}{{\cal O}}

\newcommand{\av}{{\vec{a}}}
\newcommand{\omegav}{{\vec{\omega}}}
\newcommand{\etav}{{\vec{\cal I}}}
\newcommand\mean[1]{\langle#1\rangle}
\newcommand\mmean[1]
        {\langle\hspace{-0.06cm}\langle#1\rangle\hspace{-0.06cm}\rangle}

\newcommand\bp{{\mathbf p}}
\newcommand\bW{{\mathbf W}}
\newcommand\bx{\nabla_{\mathbf s} {\cal H}}
\newcommand\bm{{\mathbb M}}

\begin{document}

\title
[Correlation of agents in a simple market:
a statistical physics perspective]
{Correlated adaptation of agents in a simple market:
a statistical physics perspective}

\author{J. P. Garrahan, E. Moro and D. Sherrington}

\address{Theoretical Physics, University of Oxford, 
1 Keble Road, Oxford OX1 3NP, United Kingdom}

\begin{abstract}
We discuss recent work in the study of a simple model for the
collective behaviour of diverse speculative agents in an idealized
stockmarket, considered from the perspective of the statistical
physics of many-body systems. The only information about other agents
available to any one is the total trade at time steps. Evidence is
presented for correlated adaptation and phase transitions/crossovers
in the global volatility of the system as a function of appropriate
information scaling dimension. Stochastically controlled irrationally
of individual agents is shown to be globally advantageous. We describe
the derivation of the underlying effective stochastic differential
equations which govern the dynamics, and make an interpretation of the
results from the point of view of the statistical physics of
disordered systems.
\end{abstract}

\section{Introduction}
There is currently much interest in the physics community in complex
co-operative behaviour of systems of many individual entities
influencing one another competitively. In particular, when combined
with non-uniformity in the inclinations of the individuals, the
behaviour of the whole can exhibit much greater complexity, richness
and subtlety than is present in the rules governing individuals; in
the words of P.\ W.\ Anderson ``more is different''
\cite{anderson}. Examples are found in spin glasses (disordered
magnetic alloys), neural networks and hard optimization problems (for
recent reviews see \cite{young,sherrington99}). Economic markets
also involve many individuals whose desires are not all simultaneously
satisfiable and who often have different inclinations and strategies.
It is therefore natural to ask to what extent the problems in physics
and in economics are similar or different, to what extent the
techniques and concepts developed for the physics problems can be
applied to those in economics, to what extent the economics problems
pose new challenges for the physicists and to search for fruitful
symbiosis of understanding, quantification and application.  To this
end we discuss in this paper recent developments in the study of a
model inspired by economics, but analyzed from the perspective of
physics, finding unexpected new results and subtleties, and concluding
that both subjects have something to teach each other and that there is
potential for further transfers and discoveries. We make no attempt
to be encyclopaedic or chronologically historical.

Before discussing the specific model, some general remarks are
appropriate.  Much of the progress in physics has come from starting
with the simplest but non-trivial microscopic entities and interaction
rules which can still lead to complex behaviour at the
macroscopic\footnote[1]{By `macroscopic' we refer to quantities which are
averaged over the behaviour of all the `microscopic' individuals.}
level. Greater ``reality'' at the microscopic level can be added
later. Many results are robust to the microscopic details, although
new features can also arise with sufficient qualitative change. We
apply a similar philosophy here, deliberately oversimplifying at the
individual level to expose novel consequences of cooperation
uncluttered by microscopic complication.  Thus, we concentrate on
systems with simple microscopic dynamical rules and minimal number of
control parameters. In the spirit of statistical many-body theory we
concentrate on systems with many ($N \gg 1$) microscopic players, with
particular regard to the leading large-$N$ behaviour of macroscopic
quantities.  We allow for temporally-fixed variation among individuals
but, in the spirit of statistical relevance, we draw their individual
characteristics independently from identical distributions.  We also
allow for stochasticity (temporal indeterminacy) at the individual
operational level, but again in a statistically relevant and minimal
parameter fashion.  As usual in statistical physics, we expect
self-averaging\footnote{By ``self-averaging'' we mean that, in the
limit $N \to \infty$ the value in a typical realization is the same as
the average over realizations.} of normal macroscopic observables,
although non-self-averaging might be envisaged at a more sophisticated
level \cite{mezard87,sherrington99}.

It is also appropriate to contrast our study with those of
conventional economics theory.  A typical assumption used in
neoclassical economic theory \cite{econo} --especially game theory
\cite{games}-- is that agents are hyper-rational.  They know the
utility functions of other agents, they are fully aware of the process
they are embedded in, they make optimum long-run plans, and so forth.
This is a rather extravagant and implausible model of human behaviour,
especially in situations like a stock market.  Moreover in
neoclassical economic theory microscopic equilibrium is the reigning
paradigm \cite{econo}.  Individual strategies are assumed to be
optimal given expectations, and expectations are assumed to be
justified given the evidence.  Equilibrium is thus reached in one-step
dynamics once hyper-rationality is assumed.  In this paper we consider
a different, perhaps more realistic, scenario in which the only
information available to any agent about the others is of the
macroscopic consequences of the multiplicity of their actions (i.e.
the analogues of market indices).  We allow for diversity and
irrationality in that they do not all draw the same conclusions from
this information \cite{Arthur94}, nor do they necessarily operate
deterministically.  In general there will not be microscopic
equilibrium although there may be macroscopic equilibrium.\footnote{By
`microscopic equilibrium' we mean a situation in which it is
disadvantageous for any individual to change his state, whereas
`macroscopic equilibrium' refers to a situation in which the
thermodynamically relevant (leading $N$) macroscopic observables do
not change with time, even though individual microscopic states do
change.}

The paper is organized as follows. In section 2 we present the
Minority Game \cite{challet1} and review its main features. This model
is a specific realization of Arthur's ``El Farol'' Bar Problem
\cite{Arthur94}, and is the starting point of our investigations. In
section 3 we consider a continuous generalization of the model, and
study the effect of allowing for stochastic decision making on the
part of the agents.  The derivation of a fundamental analytic theory
is discussed in section 4, where the underlying stochastic
differential equations for the dynamics of the system are presented.
Section 5 contains an interpretation of the results and we give our
conclusions in section 6.

\section{The Minority Game}

The model system we consider is one known as Minority Game (MG)
\cite{challet1} and is intended to mimic in a simplified way a market
of agents bidding to profit by buying when the majority
wish to sell (so that the price can be lowered) and selling
when the majority wish to buy (so that a higher price can be negotiated)
\cite{Arthur94}. It comprises a large number of agents each of whom
can act as buyer or seller, deciding on how to play at each time-step
through the application of a personal strategy to commonly
available information.\footnote[1]{A strategy is an operator
which acts on a set of data, referred to as the ``information'',
to yield an outcome which is a buy or sell instruction.}  Each 
agent has a small set of available strategies, drawn randomly,
independently and immutably with identical probabilities from a
large suite of strategies. At each time step each agent picks one of his
or her strategies, based on points allocated cumulatively to the 
strategies according to their (virtual) performance in predicting the
minority action. For simplicity no other rewards are given.  

The system has quenched (fixed in time) randomness in the set of
strategies picked at the start of the game by each agent, and it has
frustration\footnote{``Frustration'' refers to an inability to
satisfy simultaneously all the inclinations of all the microscopic
entities, and is believed to be a fundamental ingredient in producing
the complexity observed in glassy systems (where quenched disorder is
often present too) \cite{mezard87,young,sherrington99}.} in that the
rewards are for minority action, so not all the individual
inclinations can be satisfied simultaneously. There is no direct
interaction between agents. Nevertheless, correlation does arise
through the adaptive evolution of the use of strategies and manifests
itself in the interesting non-trivial macroscopic behaviour of the
system.

In the original formulation of the MG \cite{challet1} agents could
only make two choices, buy or sell, with no weight attributed to the
size of the order. Conventionally, the strategy points are set initially 
to zero and thus the agents start making their first choice at random.
The common information on which they based their
decisions was the minority choice (buy or sell) over the last $m$
time-steps. The strategies were quenched randomly-chosen Boolean
functions acting on this information, the binary output determining
the buy/sell decision. The strategy used by any agent at any time was
the one with the currently greatest point-score from those at his/her
disposal.  Numerical simulations showed \cite{savit} that while the
average in time (and over the realizations of the disorder) of the
total action was just an equality of buyers and sellers, due to the
symmetric nature of the model, the standard deviation of its
fluctuations away from this value (the analogue of the volatility of a
conventional stockmarket) displayed remarkable structure.  As a
function of the ``memory'' $m$, the volatility has two regimes: for
low values of $m$, the volatility is larger than the value
corresponding to all agents just playing randomly; it decreases
monotonically with increasing $m$, crosses the value corresponding to
random behaviour, and reaches a minimum at a critical value of the
memory $m_c$; it then starts to grow monotonically with $m$,
asymptotically approaching the random value from below.  This
non-trivial behaviour of the fluctuations was interpreted as an
indication of a cooperative ``phase transition'' in the system
\cite{savit,cm} (see, eg., \cite{julia}, for a introduction to  phase
transitions).  Simulations also showed \cite{savit} that the relevant
scaling variable was the reduced dimension of the space of strategies
$d=2^m/N$, and that the volatility scaled with the number of agents as
$\sqrt{N}$.

A second interesting observation was made in \cite{andrea}, where it
was shown numerically that that the macroscopic behaviour of the MG
was unaffected by replacing the time history by an artificial history,
chosen randomly and independently at each time-step from the space of
all possible histories with uniform probability, provided all agents
received the same bogus information (and that the point-scores were
still updated on the basis of performance).  This is a consequence of
the ergodic nature of the MG \cite{mino}, and indicates that the
principal role of the information is in providing a correlation
mechanism between the agents in terms of the strategies used. This
observation offers a great simplification for the analysis of the
model, since replacing the true history by just external noise allows
us to study a simpler system which is stochastic but local in time,
instead of the more difficult original problem which is deterministic
but non-local in time (see also \cite{relevance}). In fact, we shall
also see the consequence of allowing for a further different kind of
stochasticity in the next section.

\section{The Thermal Minority Game}

As discussed in the introduction, it is natural to expect that the
qualitative features of the MG are robust under changes in the
microscopic detail of the model.  In this section we present a
generalization of the MG to continuous degrees of freedom and to allow
for stochastic decision making on part of the agents, known as the
Thermal Minority Game (TMG), and first introduced in \cite{mino}.
This generalization not only preserves the main features of the MG
discussed so far, but also gives interesting and advantageous new
behaviour, and enables a simpler analytic description of the
coarse-grained microdynamics of the system.

\subsection{Continuous MG}

The continuous version of the MG is as follows \cite{mino}.  The
system consists again of $N$ agents playing the game.  At each time
step $t$, each agent reacts to a common piece of ``information''
$\vec{\cal I}(t)$, by taking an action or bid $b_i(t)$
($i=1,\dots,N$).  Following the observation of \cite{andrea} the
information $\vec{\cal I}(t)$ is taken to be a random noise, defined
as a unit-length vector in a $D$-dimensional space,\footnote[1]{Here $D$
is the analogue of $2^m$ in the original model.} for instance
${\mathbb R}^D$, $\delta$-correlated in time and uniformly distributed
on the unit sphere.\footnote{This is a convenient choice. Any other
normalized isotropic distribution in ${\mathbb R}^D$, e.g., a
Gaussian, would be qualitatively equally suitable. The same applies to
the strategies.}  The bid $b(t)$ is defined to be a real number, which
can be interpreted as placing an order in a market, of size $|b(t)|$
and positive/negative meaning buy/sell.  Bids are prescribed by
strategies: maps from information to bid, ${\mathbb R}^D \to {\mathbb
R}$.  For simplicity the strategy space $\Gamma$ of the model is
restricted to the subspace of homogeneous linear mappings, in contrast
to the whole space of binary functions in the MG. Thus a strategy
$\vec R$ is defined as a vector in ${\mathbb R}^D$, subject for
normalization to the constraint $\| \vec R \| = \sqrt{D}$, and the
prescribed bid is the scalar product $\vec R \cdot \vec{\cal
I}(t)$.  Each agent has $S$ strategies, drawn randomly and
independently from $\Gamma$ with uniform distribution, remaining fixed
throughout the game.  In what follows we will focus for simplicity on
the case of two strategies per agent, $S=2$, the generalization to $S >
2$ being straightforward (and the case $S=1$ being trivial, since
there is no possibility of adaptation on the part of the agents).  At
time step $t$ each agent $i$ chooses one of his/her strategies $\vec
R^\star_i(t)$ to play with.  The ``total bid'' (or ``excess demand'')
is then $A(t) \equiv \sum_{j} b_j(t) = \sum_{j} \vec R^\star_j(t)\cdot
\vec{\cal I}(t)$.  The agents keep track of the potential success of
the strategies by assigning points to them, which are updated
according to $P(\vec R,t+1) = P(\vec R,t) - A(t) \, b(\vec R) / N$,
where $P(\vec R,t)$ represents the points of strategy $\vec R$ at time
$t$.

Let us now see whether the results obtained with this continuous
formulation of the MG are the same as in the original binary
setup. To this end we first review the results of simulations. 
The average of the total bid $A(t)$ over time and quenched
disorder is zero, as is expected from the symmetric nature of the
model.  As discussed in the previous section, the first relevant
macroscopic observable is the standard deviation $\sigma$ of the total
bid, or volatility, $\sigma^2 \equiv N^{-1} \overline{\langle A^2(t)
\rangle}$, where the overline means disorder average, $\langle \cdot
\rangle \equiv \lim_{t \to \infty} \frac{1}{t}\int_{t_0}^{t_0+t}
(\cdot) \, dt'$, and we have normalized $\sigma$ by $\sqrt{N}$
according to the findings of \cite{savit}.  In Fig.\ 1 we show that
the main features of the MG are reproduced: first, the relevant
scaling parameter is the reduced dimension of the strategy space
$d=D/N$; second, the volatility starts for low $d$ at a value larger
than the one corresponding to the agents choosing randomly,
$\sigma_r=1$ in this case, decreases monotonically until it reaches a
minimum at $d=d_c(S)$, the minimum being shallower the higher is the
number of strategies $S$ \cite{challet2}, and then it approaches
$\sigma_r$ asymptotically from below. It is easy to check that {\it
all} the other standard features of the binary model are reproduced in
the continuous formulation.

\begin{figure}
\begin{center}
\leavevmode
\epsfxsize=4in
\epsffile{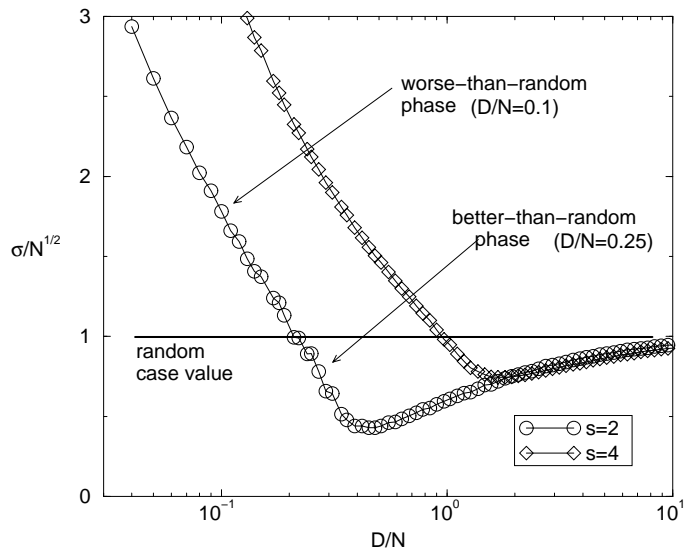}
\end{center}
\caption{Continuous formulation of the MG \cite{mino}: The scaled
volatility $\sigma/\sqrt{N}$ as a function of the reduced dimension
$D/N$, at $S=2$ and $S=4$. The horizontal line is the variance in the
random case. The total time $t$ and the initial time $t_0$ are $10000$
steps. Average over $100$ samples, $N=100$. }
\label{fig1}
\end{figure}

\subsection{Stochastic decision making}

In the original formulation of the MG the agents played in a
deterministic fashion using their `best' strategies, the ones with the
highest number of points.  In this subsection we introduce
indeterminacy (irrationality) on the part of the agents and show this
it can be advantageous.

In the TMG a natural generalization to non-deterministic behaviour
is allowed \cite{mino}.  At time step $t$, each agent $i$ chooses
$\vec R^\star_i(t)$ randomly from his/her $\{ \vec R_i^a \}$
with probabilities $\{ \pi_i^a(t) \}$ ($a = 1,\ldots,S$).  The probabilities
$\pi^a_i(t)$ are functions of the points parameterized by a
``temperature'' $T$, defined so as to interpolate between the MG case
at $T=0$, all the way up to the totally random case
$\pi_i^a=1/S$ at $T=\infty$. The temperature can be thought of as
a measure of the power of resolution of the agents: when $T=0$ they
are perfectly able to distinguish which are their best strategies,
while for increasing $T$ they are more and more confused, until for
$T=\infty$ they choose their strategy completely at random.  In the
language of Game Theory, when $T=0$ agents play `pure' strategies,
while at $T>0$ they play `mixed' ones \cite{games}.

We now consider the consequences of having introduced the temperature.
For simplicity we assume $S=2$. For the probabilities we choose the
form $\pi^{1,2}_i(t) \propto \exp{[\pm \beta \; \mbox{sgn}(p_i(t))]}$
(with $\sum_a \pi^a_i(t)=1$ and $\beta=1/T$) \cite{comino}, which
satisfies the requirements of the previous paragraph.  Consider now a
value of $d$ belonging to the worse-than-random region of the MG (see
Fig.1) and let us see what happens to the volatility $\sigma$ when we
switch on the temperature. We know that for $T=0$ we must recover the
same value as in the ordinary MG, while for $T\to\infty$ we expect to
obtain the value $\sigma_r$ of random choice. But in between a very
interesting thing occurs: $\sigma(T)$ is not a monotonically
decreasing function of $T$, but there is a large intermediate
temperature regime where $\sigma$ is {\it smaller} than the random
value $\sigma_r$; see Fig.2.  The meaning of this result is the
following: even if the system is in a MG phase which is worse than
random, there is a way to significantly decrease the volatility
$\sigma$ below the random value $\sigma_r$ by {\it not} always using
the best strategy, but rather allowing a certain degree of individual
error.
\begin{figure}
\begin{center}
\leavevmode
\epsfxsize=4in
\epsffile{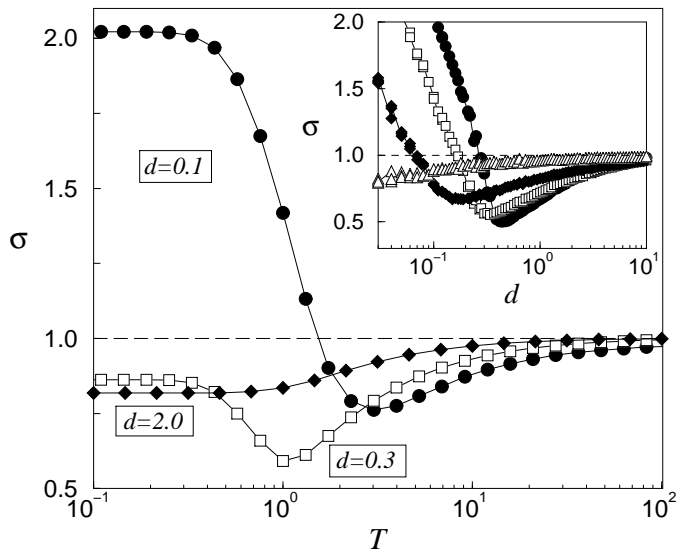}
\end{center}
\caption{TMG: Volatility as a function of the temperature
for the case of probabilities 
$\pi^{1,2}_i(t) \propto \exp{[\pm \beta \; \mbox{sgn}(p_i(t))]}$. Inset:
volatility as a function of $d$ for different values of the
temperature $T=10^{-3}, \, 1, \, 2, \, 10$.}
\label{fig2}
\end{figure}
Furthermore, even if we fix $d$ at a value belonging to the
better-than-random region, but with $d<d_c$, a similar range of
temperature still improves the behaviour of the system, decreasing the
volatility even below the MG value (see Fig.\ \ref{fig2}). In the
phase $d>d_c$ the behaviour changes; the optimal value of $\sigma$ is
at $T=0$, and the volatility simply increases with increasing
temperature towards $\sigma_r$, as shown in Fig.\ \ref{fig2}.

\begin{figure}
\begin{center}
\leavevmode
\epsfxsize=3.3in
\epsffile{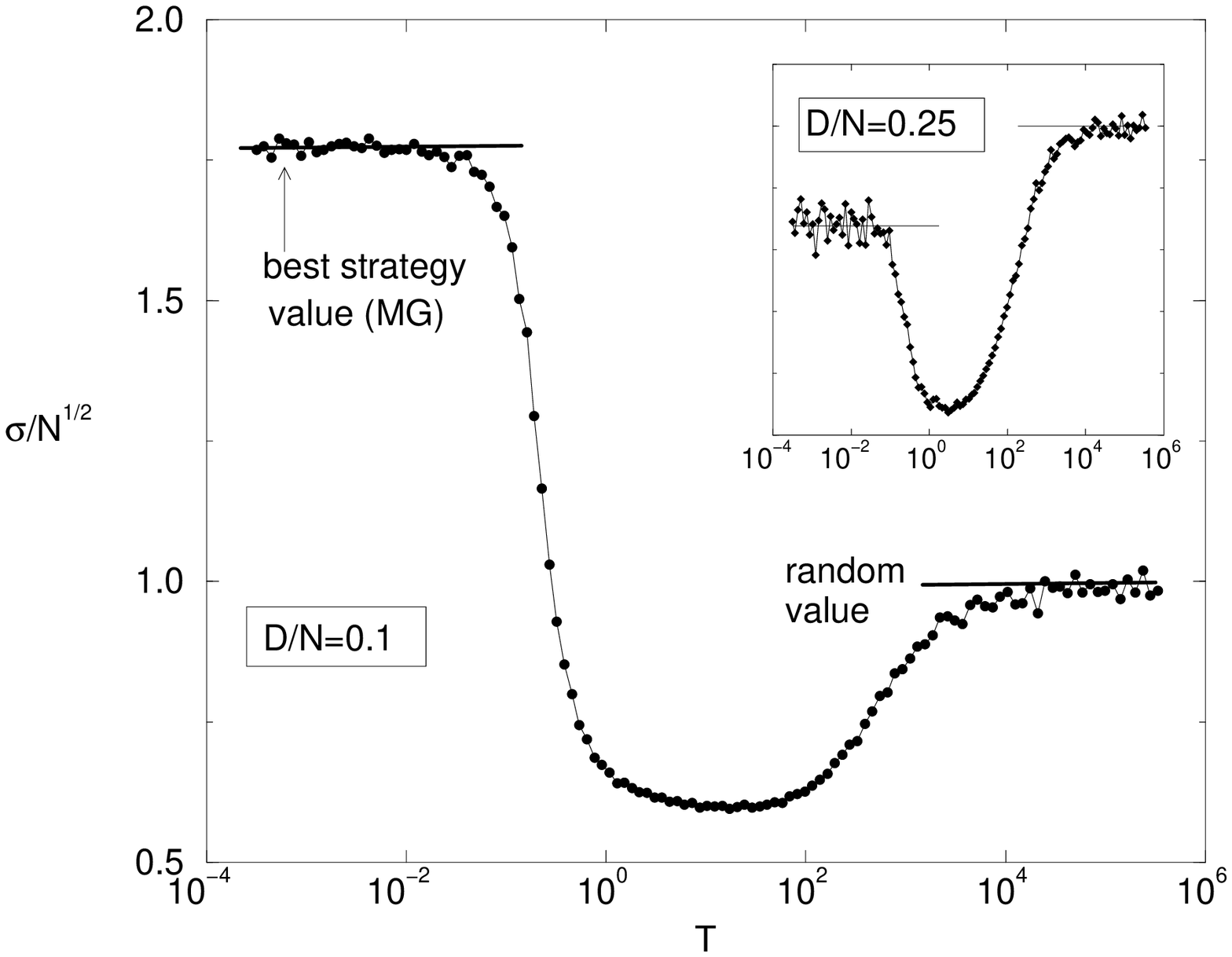}
\epsfxsize=3.3in
\epsffile{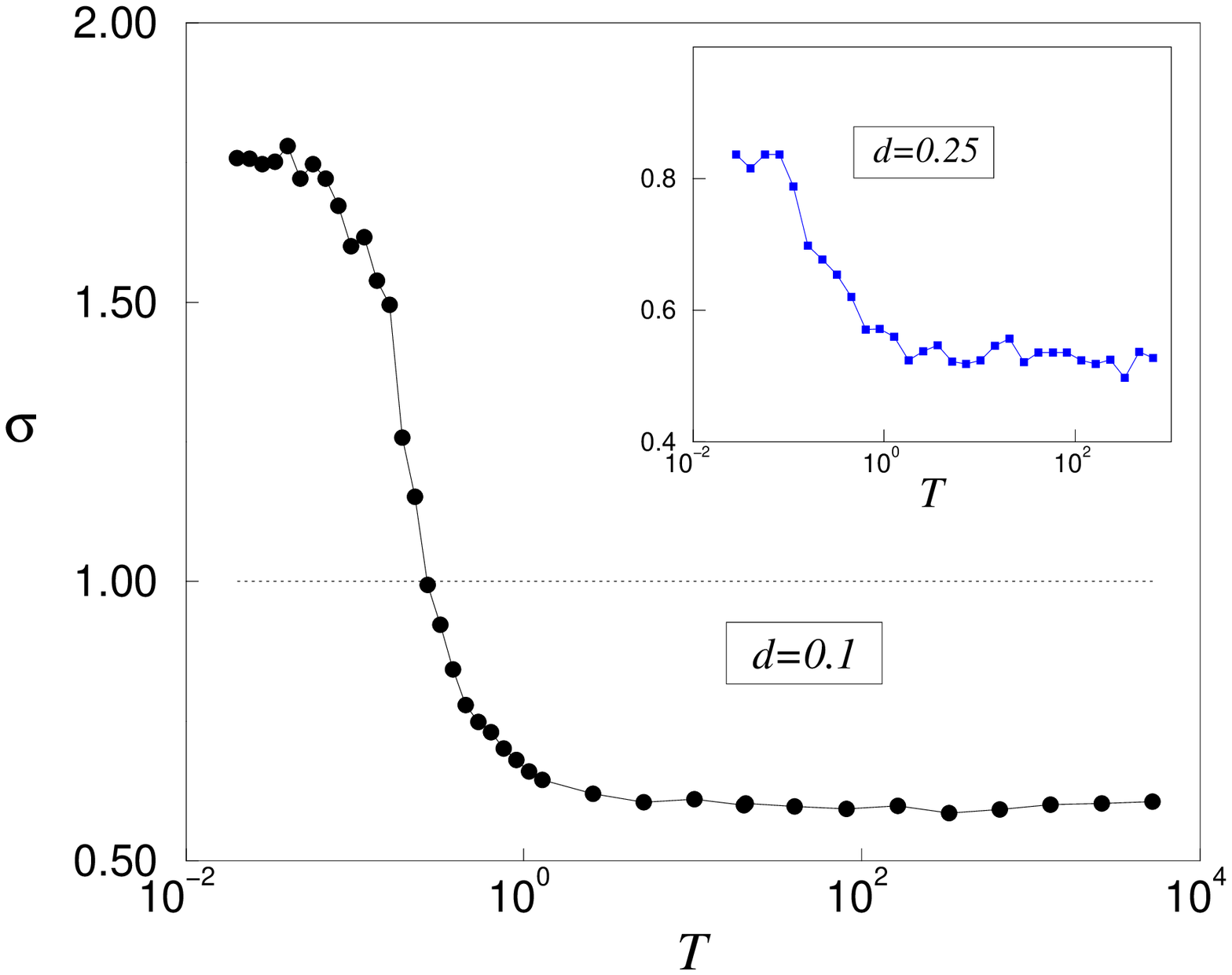}
\end{center}
\caption{(a) TMG: Volatility as a function of the temperature $T$ at
$d=0.1$, for the case $\pi^a_i(t) \propto \exp{[\beta P(\vec
R_i^a,t)]}$, for finite waiting time.  In the inset we show $d =
0.25$.(b) The same as in (a) for waiting times larger than $N T$.}
\label{fig3}
\end{figure}

Another possible functional form for the probabilities is $\pi^a_i(t)
\propto \exp{[\beta P(\vec R_i^a,t)]}$.  For long but finite
simulation time this yields the results shown in Fig.\ 3a \cite{mino},
which are analogous to those of Fig.\ 2.  However, the upturn of
$\sigma$ for large temperatures for this choice of probabilities is
only a transient \cite{bdd,comment,reply}: for any $T \gg 1$, if one
waits for times of order $N T$ the volatility stays at its smallest
possible value; see Fig.\ 3b. Thus, we see that in the $d <d_c$ phase,
for any finite temperature the performance of the system will be
better than the original MG, and for {\em any finite} temperature $T >
\Ocal(1)$ the performance will be optimal, provided one waits long
enough.

\section{Continuous time dynamics of the TMG}

In this section we derive the continuous time limit of the TMG as a
starting point for the analytical study of its dynamics
\cite{comino}. We do this in two steps. We first show that, to a good
approximation, the external information can be eliminated in favour of
an effective interaction between the agents. We then cast the
dynamical equations of the TMG as a set of stochastic differential
equations for an interacting disordered system with nontrivial random
diffusion. Again, for simplicity, we consider explicitly $S=2$.

The set of unconstrained degrees of freedom of the TMG is given by the
differences $\{ p_i(t) \}$ of the points of the two strategies of each
agent.  The choice of strategies used at each stage is given by $\vec
R^\star_i(t) = \vec \omega_i + \vec \xi_i \, \mbox{sgn} \left[ s_i(t) + \mu(t)
\right]$, where $\vec
\omega_i \equiv \left( \Rv_i^1 + \Rv_i^2 \right)/2$, $\vec \xi_i
\equiv \left( \Rv_i^1 - \Rv_i^2 \right)/2$, $s_i(t) \equiv \pi_i^1(t)
 - \pi_i^2(t)$, and $\mu(t)$ is a
stochastic random variable uniformly distributed between $-1$ and $1$
and independently distributed in time.  The equations for the point
differences then read,
\beq
p_i(t+1) = p_i(t) - \av(t) \cdot \etav(t) \, 
    \vec \xi_i \cdot \etav(t)
    \label{pipi} ,
\eeq
where $\av(t) \equiv \sum_{i} \vec R^\star_i(t)/N$.  Eqs.\
(\ref{pipi}), together with the random processes for $\etav(t)$ and
$\Rv_i^\star(t)$, define the dynamics of the TMG.

We now consider the continuous time limit of
Eqs.\ (\ref{pipi}) in such a way as to preserve all the macroscopic 
features of the TMG. 
To this end we introduce an arbitrary time step $\Delta t$.
We deal first with the information $\etav(t)$. 
Let us assume that $\etav(t)$ is a differential random motion
in the space of strategies, i.e.,
$\etav(t) = \Delta \vec W(t)$,
with zero mean and variance $\Delta t$. 
Replacing in Eqs.\ (\ref{pipi}) we obtain
$p_i(t+\Delta t) = p_i(t) - 
    \av(t) \cdot \Delta \vec W(t) \,
    \xiv_i \cdot \Delta \vec W(t)$. In the
limit $\Delta t \to 0$, and using the 
Kramers-Moyal expansion \cite{vankampen}, 
we get
\beq\label{eqcont_1}
d p_i(t) = -\frac{1}{ND} \sum_{j} \vec R^\star_j(t)\cdot \xiv_i \, dt
    + \Ocal(dt^2) .
\eeq
Note that to $\Ocal(dt)$ the noise has been eliminated
in favour of an effective strategy interaction among the agents, and
the standard deviation becomes
\begin{equation}\label{sigma2} 
\sigma^2 = \frac{1}{ND} 
\overline{\bigg\langle \sum_{ij} \vec R^\star_i(t) \cdot \vec R^\star_j(t)
 \bigg\rangle}.
	\label{ss}
\end{equation}

When the temperature is different from zero the TMG Eqs.\
(\ref{eqcont_1}) still depend on the stochastic choice of strategies
$\Rv_i^\star(t)$, even at leading order.  At each time step, $\Rv_i^*$
takes one of the two possible values $\Rv_i^{1,2}$, defining a
stochastic jump process. In order to write the corresponding Master
Equation we need to know the transition probabilities.  The r.h.s. of
Eq.\ (\ref{eqcont_1}), which we denote $\Delta_i$, is a normalized sum
of $N$ random numbers $\xiv_i \cdot \Rv_j^\star(t)$, each with mean
$m_{ij} = \xiv_i \cdot \omegav_j + \xiv_i \cdot \xiv_j \, s_j(t)$, and
variance $v_{ij}=( \xiv_i \cdot \xiv_j )^2 [ 1 - s_j^2(t)]$.  By the
central limit theorem, we know that for $N$ large $\Delta_i$ will tend
to be normally distributed with mean $\mmean{\Delta_i} = \sum_j
m_{ij}$, and variance $\mmean{\Delta_i^2} = \sum_j v_{ij}$, 
where $\mmean{\cdot}$ stands for average over realizations
of the random process $\mu(t)$.  Moreover, $\Delta_i$ and $\Delta_{j
\neq i}$ are correlated, the covariance matrix given by 
\beq
M_{ij}[\bp(t)] \equiv \mmean{\Delta_i \, \Delta_j} - \mmean{\Delta_i}
\mmean{\Delta_j} = \sum_k \xiv_i \cdot \xiv_k \, \xiv_j \cdot \xiv_k
\left[ 1 - s_k^2(t) \right] , \label{corr} 
\eeq 
where $\bp \equiv
(p_1, \ldots, p_N)$, etc.  Collecting these results, we obtain the
transition probabilities in the large $N$ limit, $W(\bp'| \bp) =
\Phi(\bx; \bm)$, where $\Phi$ corresponds to the normal distribution
with mean $\bx$ and covariance matrix $\bm \equiv \{ M_{ij} \}$,
where the ``Hamiltonian'' $\cal H$ is given by\footnote[1]{This 
expression was
first obtained, by a different procedure and with a different
interpretation, in \cite{cm}.}
\beq 
{\cal H}= \frac{1}{2} \Omega + 
        \sum_i h_i s_i + 
        \frac{1}{2} \sum_{ij} J_{ij} s_i s_j ,
	\label{H}
\eeq
with
\beq 
h_i \equiv \sum_j \omegav_j \cdot \xiv_i / ND
	, \;\;\; 
	J_{ij} \equiv \xiv_j\cdot\xiv_i /ND .
	\label{hJ}
\eeq 
Note
that $\partial {\cal H}/\partial s_i \sim \Ocal(1)$, and $M_{ij} \sim
\Ocal(1/N)$, so that fluctuations are also of $\Ocal(1)$ and thus are
{\em not} suppressed when $N \to \infty$.

The $\mu(t)$ are chosen independently at 
each time. If we make the natural assumption 
that in the limit $dt \to 0$ their 
correlation at different times is a $\delta$-function, 
the Master Equation becomes a Fokker-Planck equation 
by means of Kramers-Moyal expansion \cite{vankampen}
\beq\label{pdf1}
\frac{\partial {\cal P}}{\partial t}
        = 
        - \sum_i \frac{\partial}{\partial p_i}
        \left( \frac{\partial{\cal H}}{\partial s_i} 
        \, {\cal P} \right) 
        + \frac{1}{2} \sum_{ij}
        \frac{\partial^2}{\partial p_i \partial p_j}
        \left( M_{ij} \, {\cal P} \right)  .
\eeq

The dynamics of the TMG 
is therefore effectively described by
a set of stochastic differential equations for the point
differences
\beq
d\bp = - \bx \, dt +    
        \bm \cdot d\bW ,
        \label{sde}
\eeq
where $\bW(t)$ is an $N$-dimensional Wiener process, 
and the volatility is given by
$\sigma^2 = 2
        \overline{ \mean{{\cal H}}} 
        + \sum_i \overline{J_{ii}}
        - \sum_i \overline{J_{ii} \mean{s_i^2}}$.

A detailed interpretatin of ${\cal H}$ is given in the next section,
but briefly eq. (\ref{sde}) is suggestive of it as controlling
energy with the `motion' of $\bp$ given by its derivative. However we note
that for a natural analogue of Newton's law or its generalization to 
a noisy environment, as in the Langevin equation, the derivative
would be with respect to $\bp$, whereas here it is with respect to 
${\mathbf s}$ (which is a function of $\bp$), so that a metric is needed
to relate $d\bp/dt$ to the natural force $\bx$. At finite temperature
one also has the unusual extra diffusive/noise term $\bm \cdot d\bW$. An
investigation based on replacing $\bx$ by $\nabla_{\bp} {\cal H}$
and ingnoring the diffusive term was performed in 
\cite{cmz}.
 
\section{Interpretations}

\begin{figure}
\begin{center}
\leavevmode
\epsfxsize=4in
\epsffile{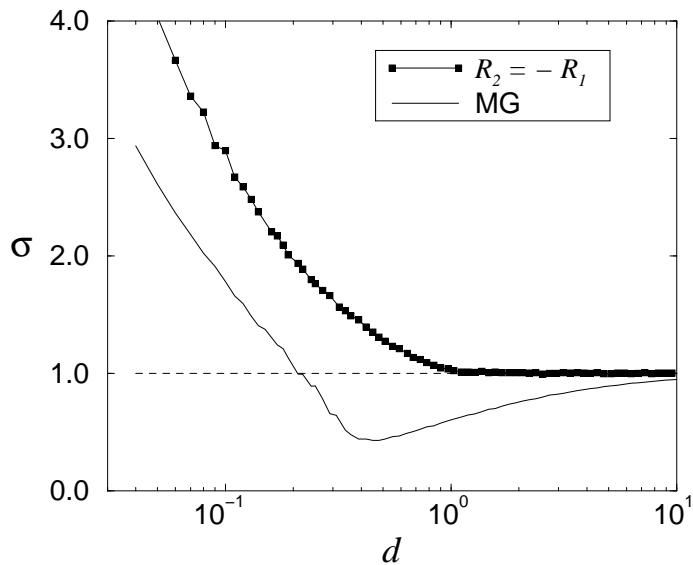}
\end{center}
\caption{Volatility for anticorrelated strategies
$\vec{R}^2_i=-\vec{R}_i^1$ (squares) compared with the standard
MG (solid line).}\label{fig4}
\end{figure}

Notwithstanding the subtleties mentioned in the last paragraphs, it is
interesting to consider the implications of ${\cal{H}}$ as a
controlling function of the dynamics. The form of ${\cal{H}}$
exhibited in Eq.\ (\ref{H}) is a familiar one in statistical physics,
with $- J_{ij}$ interpreted as measuring the strength of correlation
between spins $s_i$ and $s_j$ and $- h_i$ as a (magnetic) field acting
on $s_i$ and trying to ``orient'' it. Hence is formally justified the
interpretation of common `information', to which all respond, as
providing a mechanism of effective mutual interaction\footnote[1]{In
fact, in conventional statistical physics, the converse is often
employed in formal analysis, replacing direct interaction by
interaction through randomly-distributed global intermediates
\cite{Hubbard59,Sherrington67,Sherrington71}.}. More
particularly, taking account of the random character of $\{h_i\}$ and
$\{J_{ij}\}$, ${\cal{H}}$ is reminiscent of the
Sherrington-Kirkpatrick model \cite{sk} of a spin glass and the
Hopfield model \cite{hop} of a neural network, in both cases
augmented by random fields \cite{young}. In the SK model the $J_{ij}$
are chosen randomly from a distribution with variance scaling as
$N^{-1}$, while for the Hopfield model $J_{ij}$ is as given in Eq.\
(\ref{hJ}), but with opposite sign and where the Cartesian components
of $\{\xiv_i\}$ correspond to memorized patterns of activity.  In both
of these models (SK and Hopfield) the ground state energy (minimum of
the corresponding ${\cal{H}}$) is less than the energy associated with
a paramagnet (value of ${\cal{H}}$ corresponding to random $\{s_i\}$)
due to judicious correlation of the $\{J_{ij}\}$ and $\{s_i\}$, even
with all the $\{h_i\} =0$. It is then natural to ask whether a similar
correlation is responsible for the reduction of the volatility of the
MG below that for random operation.  This is however not the case: in
Fig.\ \ref{fig4} \cite{kcl} we show the result of numerical simulations of a system in
which the second strategy of each agent $\Rv^2_i$ is exactly the
opposite of its first strategy, that is $\Rv^2_i = -\Rv^1_i$ (with
$\Rv^1_i$ still chosen randomly at the start of the game), so that all
the $h_i$ are zero, and we note that $\sigma$ never falls below the
random value.  Clearly, the random field term is crucial in reducing
the volatility advantageously \cite{cm,cmzhang}.

\begin{figure}
\begin{center}
\leavevmode
\epsfxsize=4in
\epsffile{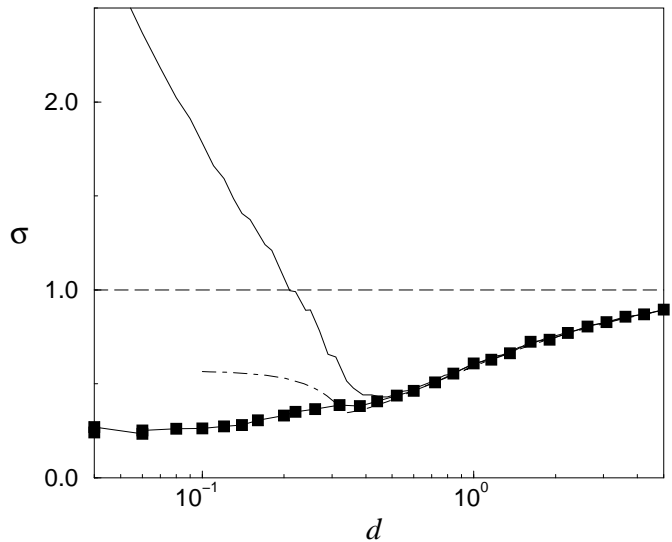}
\end{center}
\caption{Volatility as a function of $d$ for
random initial conditions (squares) compared with the standard MG
(solid line). Dotted line corresponds to the
approximation of \cite{cmz}.}\label{fig5}
\end{figure}

The first term in the dynamical equations (\ref{sde}) corresponds (up
to a factor) to gradient descent in the surface defined by ${\cal H}$.
The natural question then is whether the asymptotic states reached by
the dynamics are given by the typical extrema of ${\cal H}$, which
would imply that in the long run the system reaches macroscopic
equilibrium. This issue was explored in \cite{cmz} (see also
\cite{mcz}), where ${\cal H}$ was minimized employing techniques
developed for the statistical mechanics of spin glasses
\cite{Edwards75,sk} and adapted for neural networks \cite{Amit85} (for
reviews see \cite{mezard87,sherrington93,sherrington99}).  An
excellent agreement with the numerics of the MG was found in the phase
$d>d_c$, but this method failed to reproduce the behaviour in the
$d<d_c$ phase (see Fig.\ref{fig5}), although an equilibrium transition was
found at a value of $d$ close to $d_c$.\footnote{In the case of the
TMG with probabilities $\pi^a_i(t) \propto \exp{[\beta P(\vec
R_i^a,t)]}$, the asymtotic value of the volatility for large $T$ (see
Fig.\ 3b) coincides with the one predicted from the equilibrium
calculation of \cite{cmz} (see also \cite{comment}). This can be
understood from Eqs.\ (\ref{sde}): for large $T$, a systematic
rescaling of time and points can be used to eliminate the diffusive
term in the equations, and the effective dynamics becomes independent
of $T$ \cite{comino}. Note that this is a consequence of the
functional form of the probabilities, and does not hold in the case
$\pi^{1,2}_i(t) \propto \exp{[\pm \beta \; \mbox{sgn}(p_i(t))]}$.}

This seems to suggest that the behaviour of the MG in the $d<d_c$
phase is dynamical in nature. A simple test is the sensibility to
initial conditions.  In Fig.\ \ref{fig5} we show the results of
simulating the original dynamics Eqs.\ (\ref{pipi}) of the MG starting
from any initial conditions with $|p_i(0)| \sim \Ocal(1)$
\cite{comino}, instead of the conventional choice of $p_i(0) =
0$.\footnote[1]{Note that with the conventional choice the agents do
not prefer initially any of their strategies, while when $|p_i(0)|
\sim \Ocal(1)$ agents have initially a preferred strategy.}  We can
see that the the behaviour of the system is very different in the
region $d<d_c$: after an initial transient, the variance falls below
the initial random value and stays in the better-than-random phase for
all values of $d$. This sensitivity of the results to the initial
conditions is a clear indication that the system does not equilibrate
for $d<d_c$. An important open problem in the MG is finding the
relation between the equilibrium (static) phase transition obtained
from minimization of the Hamiltonian ${\cal H}$ \cite{cmz} and the
actual dynamical transition observed in the
simulations.\footnote{In the case of anticorrelated strategies of
Fig.\ \ref{fig4}, minimization of ${\cal H}$ gives an equilibrium
transition at $d=1$ \cite{cmzhang} while the $d<1$ phase is
out-of-equilibrium.}

The differential equations (\ref{pdf1}) and (\ref{sde}) provide the 
basis for development of a macroscopic dynamics, either via the
dynamical replica theory of Coolen and Sherrington \cite{coolen} or
via the generating functional approach \cite{msr}. However, we defer
this discussion to later papers.

Finally we remark on the relationship with the crowd-anticrowd concept
of \cite{neil}, where a crowd is a group of agents playing the same
strategy and the corresponding anticrowd play the opposite one. The
proposal of \cite{neil} is that the macroscopic properties of the MG
can be described by the behaviour of the crowd-anticrowd pairs. In
this approach the volatility of the system is approximated as the sum
over all the pairs of crowd-anticrowds of the square of the difference
of their sizes. We can formalize this concept in the continuous
formulation of the MG by defining the crowd-anticrowd pairs in terms
of the projection of the strategies used by the agents on an arbitrary
orthonormal basis $\{ {\hat e}_{\mu} \} \; (\mu=1,\ldots,D)$ of the
strategy space: $N_\mu - N_{\bar \mu} \equiv \sum_i {\hat e}_{\mu}
\cdot \Rv^\star_i$.  This definition is analogous to that of ``staggered''
magnetizations in a spin system or a neural network \cite{Amit85}. If
we write the volatility using the approximation of \cite{neil}
$\sigma^2 = \frac{1}{N} \sum_\mu (N_\mu - N_{\bar \mu})^2$, and make
use of the completeness relation of the $\{ {\hat e}_{\mu} \}$, we
recover Eq.\ (\ref{ss}). Moreover, an effective dynamics of
crowds-anticrowds, as proposed in \cite{dycw}, is exactly derivable
from the microscopic dynamical equations (\ref{sde}).

It is of course interesting to aks about the stability of our conclusions
to minor perturbations of the model. We have already noted the 
stability of the qualitative features of the original minority game
to a change in the number of strategies per agent (excluding the 
special case $S=1$). This continues to the thermal extension. We have
not investigated explicitly changes to the learning rule but it
seems reasonable to expect qualitatively analogous behaviour for other
generalized minority rules, for example non-linear rewards but 
still favouring minority action \cite{savit2}. Reward rules involving
capital accumulation and consequent variation of potential market
influence \cite{farmer2} are another natural extension, as also
evolution of the strategies themselves \cite{kinzel,johnson2}

\section{Conclusion}

In this paper we have discussed the application of techniques and
philosophy of the statistical physics of complex cooperative
frustrated many-body systems to a simple model of agents in a
competitive market.  We have shown that this can lead to both
qualitatively and quantitatively new results in the economics-inspired
model and also, conversely, that economics models can yield new
challenges for statistical physics.  The techniques have included
computer simulation and analysis.  In the simulations we have
concentrated on the macroscopic steady-state.  In the analysis we have
derived a potentially useful microscopic formulation in terms of
stochastic differential equations, itself different and more subtle
that that normally encountered in conventional condensed-matter
statistical physics at the corresponding coarse-grained microscopic
level. The challenge still remains to develop a full macrodynamics,
both equilibrium and out-of-equilibrium, but the derived microdynamics
is the necessary ingredient for extension of relevant techniques from
statistical physics.

We have restricted discussion to a model which is clearly
oversimplified from the perspective of a true economic market, but it
is possible to envisage still simplified but more realistic models
which should be capable of more truly analysing the meaning of
``efficiency'' and going beyond it.

\section*{Acknowledgments}

It is a pleasure to thank Andrea Cavagna and Irene Giardina for
collaboration in part of the work reviewed in this paper.  This work
was supported by EC Grant No.\ ARG/B7-3011/94/27 and EPSRC Grant No.\
GR/M04426. We have also benefited from ESF programme SPHINX.

\end{document}